\documentclass[journal]{IEEEtran}
\makeatletter
\renewcommand\footnoterule{%
  \kern-3\p@
  \hrule\@width.4\columnwidth
  \kern2.6\p@}
  \makeatother

\usepackage{cite}
\usepackage{hyperref}
\ifCLASSINFOpdf
\usepackage[pdftex]{graphicx}
\usepackage[utf8]{inputenc}
\usepackage[english]{babel}
\graphicspath{{../pdf/}{../jpeg/}}

\else
\usepackage[dvips]{graphicx}
 \graphicspath{{../eps/}}
\fi
\usepackage{amssymb}
\usepackage[cmex10]{amsmath}
\usepackage{mathtools}

\usepackage{algorithm}  
\usepackage{algorithmic}

\usepackage{caption}

\usepackage{algorithmic}
\usepackage{array}
\usepackage{stfloats}

\usepackage{algorithm}
\usepackage{algorithmic}
\usepackage{multirow}
\usepackage{booktabs}
\usepackage{subfigure}
\usepackage{float}
\usepackage{amsmath}
\usepackage{authblk}
\usepackage[T1]{fontenc}

\begin{document}

\title{DHPrep: Deep Hawkes Process based Dynamic Network Representation}

\author{\IEEEauthorblockN{Ruixuan Han\IEEEauthorrefmark{1}, Hongxiang Li\IEEEauthorrefmark{1}, Bin Xie\IEEEauthorrefmark{2}}

\IEEEauthorblockA{\IEEEauthorrefmark{1}Department of Electrical and Computer Engineering, University of Louisville, Louisville, KY}

\IEEEauthorblockA{\IEEEauthorrefmark{2}InfoBeyond Technology LLC, Louisville, KY}
}

\pagestyle{empty}  
\thispagestyle{empty} 

\maketitle

\begin{abstract}

Networks representation aims to encode vertices into a low-dimensional space, while preserving the original network structures and properties. Most existing methods focus on static network structure without considering temporal dynamics. However, in real world, most networks (e.g., social and biological networks) are dynamic in nature and are constantly evolving over time. Such temporal dynamics are critical in representations learning, especially for predicting dynamic networks behaviors. To this end, a Deep Hawkes Process based Dynamic Networks Representation algorithm (DHPrep) is proposed in this paper, which is capable of capturing temporal dynamics of dynamic networks. Specifically, DHPrep incorporates both structural information and temporal dynamics to learn vertices representations that can model the edge formation process for a vertex pair, where the structural information is used to capture the historical impact from their neighborhood, and the temporal dynamics utilize this historical information and apply Hawkes point process to model the edges formation process. Moreover, a temporal smoother is further imposed to ensure the representations evolve smoothly over time. To evaluate the effectiveness of DHPrep, extensive experiments are carried out using four real-world datasets. Experimental results reveal that our DHPrep algorithm outperforms state-of-the-art baseline methods in various tasks including link prediction and vertices recommendation.


\end{abstract}




\IEEEpeerreviewmaketitle

\section{Introduction} Graphs or networks are widely used to describe complex systems in various applications, such as social networks \cite{huo2018link}, bioinformatics \cite{grover2016node2vec}, natural language processing \cite{perozzi2014deepwalk}, and relational knowledge bases \cite{wang2014knowledge}. In these applications, how to mine useful information from structured graph data has drawn significant research efforts in both academia and industry. In particular, one fundamental question is how to learn informative vertices representations to express networks, which essentially is to encode a network into a low-dimensional space where each vertex can be presented as a single point in the learned latent space. Such representations can facilitate various network mining tasks including information retrieval, link prediction and vertices classification  \cite{li2018deep, bhagat2011node}. 

Most existing representation learning approaches, e.g., LINE\cite{tang2015line}, Node2vec\cite{grover2016node2vec}, SDNE\cite{wang2016structural} and GraphSage\cite{hamilton2017inductive}, primarily focus on static graphs with a fixed set of vertices and edges. However, many real-world networks are dynamic and are constantly evolving over time. Examples include financial transaction networks where enterprises may switch their business partners over time, and email communication networks whose interaction structures may change due to internal affairs. Therefore, it is crucial that the learned representations for dynamic networks can capture the temporal dynamics to accurately predict vertex characteristics and future links. In the literature, existing techniques in dynamics network representations can be divided into two categories: i.) Continuous-time approaches \cite{nguyen2018continuous,trivedi2017know, trivedi2018dyrep}, where the dynamic network is observed as an aggregated network with timestamped edges. These approaches only generate one latent representation for each vertex such that the temporal evolution pattern is captured at a very coarse level and the evolving pattern cannot be explicitly revealed. ii.) Discrete-time approaches \cite{zhu2016scalable,zhou2018dynamic,li2014deep}, where the evolution of a dynamic network is observed as a collection of network snapshots over time and multiple network representations are observed in order to reveal the network evolution. While existing approaches have demonstrated their capability of capturing network dynamics, they tend to preserve (encode) very limited temporal information, causing information loss between snapshots. For example, most existing works only use the dynamics between adjacent snapshots to update representations, ignoring the long-term temporal dynamics of the networks. In fact, it is still an open problem to model and learn informative representations that capture comprehensive temporal dynamics in discrete dynamic networks.


\begin{figure}
\centering
\includegraphics[width=3.5in]{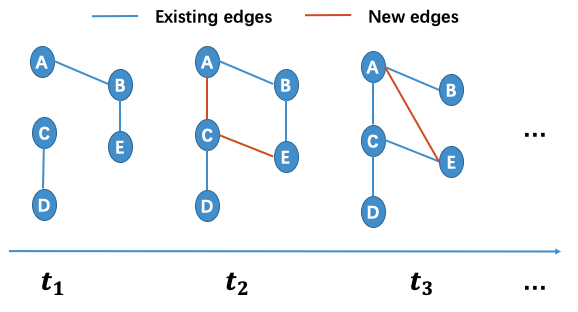}
\caption{The Illustration of dynamic networks.}
\captionsetup{font={large}}
\end{figure}

The dynamic structure of a network is essentially driven by edges establishments between vertex pairs, and each edge can be regarded as an event \cite{lu2019temporal}. Since events in real-world are correlated, a current events can be regarded as being triggered by a serial of previous events. For example, in Figure 1 vertex $A$ and $E$ are not connected with each other at $t_1$, but then establish an edge at $t_3$. The new edge at $t_3$ could be directly related to the historical node relations at both $t_1$ and $t_2$, i.e. edge $(A, B, t_1)$, $(E, C, t_2)$ and $(A, C, t_2)$.  
Besides temporal dynamics, network structure should also be considered for representation learning. The structural information embedded in representations can reflect the relationship between vertices, which in turn is useful for modeling temporal dynamics. Therefore, both long-term temporal dynamics and network structural information shall be utilized in order to learn representations in discrete dynamic networks.

In this paper, we propose a new dynamic network representation algorithm, namely DHPrep, to learn latent vertices representations of dynamic networks. In DHPrep, we first design a deep Hawkes Process (DHP) \cite{hawkes1971spectra} to model temporal dynamics. 
Then, the structure homophily of network snapshots is considered to model the effect of historical events by utilizing the first order structural information. Moreover, in order to achieve a smooth evolving network representation, a temporal smoother is further embedded in DHPrep by minimizing the Euclidean distance of the representations of two adjacent snapshots. Finally, the DHPrep integrates DHP, structure homophily and temporal smoothness into an unified model for joint optimization. Our DHPrep framework is capable of capturing the formation process of topological structures and the evolutionary patterns of temporal networks, which are essential for accurate prediction of future network structures. 
The main contributions of this paper are summarized as follows:
\begin{itemize}
	\item We propose a novel representation learning model DHPrep for dynamic networks. DHPrep is able to map the discrete network snapshots to a low-dimensional space by exploring both temporal dynamics and network structural information.

	\item We propose a joint learning architecture to simultaneously optimize DHP, structure homophily and temporal smoothness. 
	 
	\item We conduct comprehensive experiments on real-world datasets. Experimental results show that DHPrep outperforms state-of-the-art baselines in various prediction tasks. 
\end{itemize}

The rest of this paper is organized as follows. Section II introduces related works. Then, the proposed DHPrep framework is presented in Section III, followed by experimental results and discussions in Section IV. Finally, a conclusion is drawn in section V.

\section{Related Works}
In recent years, learning representations for structured data has drawn significant attention. This section summarizes related works for both static and dynamic networks.

\textbf{Static Network Representation} Static network representation methods aim to capture the structural information of a network regardless of its temporal information. Early network representation approaches exploit spectral properties of a graph to perform dimensionality reduction \cite{belkin2002laplacian, jolliffe2002principal}. Later on, inspired by the advancements of Natural Language Processing, LINE \cite{tang2015line} and Node2vec \cite{grover2016node2vec} utilize skip-gram model to learn vertex representations by maximizing the likelihood of co-occurrence in fixed-length random walks. Recently, several neural networks based network representation approaches have been proposed. In SDNE \cite{wang2016structural}, a semi-supervised model is proposed which learns representations by preserving local and  global network structures simultaneously. Hamilton et al. \cite{hamilton2017inductive} and Veličković et al. \cite{velickovic2019deep} extend graph convolution networks and propose a general unsupervised graph representation learning framework through trainable neighborhood aggregation functions. However, all these works neglect temporal dynamics and are not appropriate for dynamic network representations.

\textbf{Dynamic Network Representation} 
Existing dynamic networks representations techniques can be divided into two categories \cite{trivedi2018dyrep}, i.e., continuous-time and discrete-time approaches. For continuous-time representation, it only generates a single latent representation for each vertex, which cannot explicitly reveals the evolving pattern of the dynamic network \cite{nguyen2018continuous, zuo2018embedding}. Therefore, this paper focuses on learning representations over discrete-time dynamic network. For discrete-time approaches, DynGEM \cite{goyal2018dyngem} considers the correlation of two adjacent networks, and explores graph-autoencoders to update the current representation from previous network representations. In \cite{zhou2018dynamic}, Zhou proposes to utilize triadic closure process to generate dynamics network representations. Additionally, TIMERS \cite{zhang2018timers} incrementally updates the embedding by using incremental Singular Value Decomposition (SVD) techniques. However, these methods cannot learn long-term dynamics and have poor performance when the network experiences significant temporal variations. Unlike these approaches, our DHPrep incorporates temporal point process with structural information and is capable of capturing all relevant historical contexts to learn dynamic network representations.

\section{Our approach: DHPrep}
In order to facilitate the discussion, we first make the following definitions:

{\bf{Definition 1}}: \emph{(Dynamic Networks)}: Dynamic networks are defined as a sequence of temporal snapshots taken with a fixed time interval. In dynamic networks, we denote a series of snapshots as $\mathcal{G} = \{G_{1}, G_{2}, ..., G_{T} \}$, where $G_t = \{V, E_t, W_t\}$ represents a undirected network at time $t$ ($t \in [1, T]$). Let $V$ denote the vertex set and $E_t$ represent the edge set that occurs during the $t^{th}$ time interval. Each edge $e_{i,j}^t$ in $E_t$ is associated with a weight $w_{i,j}^t$ ($w_{i,j}^t \in W_t$).

{\bf{Definition 2}}: \emph{(Dynamic Networks Representation)}: Given a set of consecutive snapshots $\mathcal{G} = \{G_{1}, G_{2}, ..., G_{T} \}$, dynamic networks representation aims to learn a mapping function that can illustrate the evolving dynamics among network snapshots. Mathematically, the mapping function for each time instance $t$ can be represented as $f^t: v_i \mapsto \mathbb{R}^d$, where $d$ is the dimension of learned representation for vertex $v_i$. The design of $f^t$ shall consider not only the spatial information of the current network but also the temporal correlations among snapshots to reveal the true dynamics of $\mathcal{G}$ 


\subsection{Preliminary: Hakes Process}
Temporal point processes have previously been used to model network dynamics \cite{mei2017neural}. Specifically, it assumes that historical events can influence the occurrence of the current event, and models the discrete sequential events by defining a conditional intensity function (CIF). Mathematically, CIF characterizes the event arrival rate occurring in a small time window $[t, t + \Delta t]$ given all the historical events $\mathcal{H}(t)$.

\begin{equation}
\lambda(t|\mathcal{H}(t)) = \lim_{\Delta \rightarrow 0} \frac{\mathbb{E}[N(t + \Delta t | \mathcal{H}(t))]}{\Delta t}
\end{equation}

As a well-known temporal point process, Hawkes process defines \cite{mei2017neural} the conditional intensity function $\lambda(t)$ as follows:

\begin{equation}
\tilde{\lambda}(t) = \mu(t) + \int_{t}^{-\infty} \kappa(t-s)dN(s)
\end{equation}

\noindent where $\mu(t)$ is the base intensity of a particular event, representing the spontaneous event arrival rate at time $t$; $\kappa(\cdot)$ is a kernel function that models the time decay effect of historical events on the current event (e.g., exponential function and power law kernel function). The CIF in Equation (2) clearly shows that the occurrence of current event is affected not only by the last event but also by all historical events with a temporal decay effect. Such point process is desirable to model the network evolution process for discrete dynamic networks where an edge establishment is determined by all historical neighbors of the involved vertices. 


\subsection{DHPrep Framework} 
Our DHPrep combines the spatial and structural information in dynamic networks to better capture the evolution dynamics.
Specifically, DHPrep uses Hawkes process to model the dynamics of network evolution, while the spatial information is captured by the linkage status within a single network snapshot. 

\subsubsection{\bf{Deep Hawkes Process}} As discussed before, establishing an edge between two vertices can be regarded as an event. The trigger for such an event originates not only from the two involved vertices but also from their historical neighbors. Moreover, past events affect the current event to various degrees due to the temporal decay effect. Therefore, we propose a Deep Hawkes Process to capture the evolving network dynamics. 

Mathematically, given an edge (i.e., event) $e = (i, j, t)$, its occurrence intensity can be calculated as:

\begin{equation}
\tilde{\lambda}_{i,j}(t) = \mu_{i,j}(t) + \sum_{t_h < t} \alpha_{h,i}(t_h)g_{h,i}(t_h)\kappa(t - t_h)
\end{equation}

\noindent where $\mu_{i,j}(t)$ is the base intensity between the two participating vertices $i$ and $j$, and the second term is the exciting process of vertex $i$'s historical neighbors prior to time $t$. Specifically, the base intensity $\mu_{i,j}(t)$ indicates the likelihood of participants $i$ and $j$ to establish an edge. Such a likelihood can be calculated by the similarity of the two vertices because more similar vertices are more likely to establish an edge. In DHPrep, we adopt negative squared Euclidean distance to calculate the similarity of two vertices, so the base intensity is defined as: 

\begin{equation}
\mu_{i,j}(t) = - \Vert{{\bf{u}}^{t-1}_i - {\bf{u}}^{t-1}_j} \Vert ^2
\end{equation}
where the base intensity of the current event is determined by the last snapshot representation. 

The second term in Equation (3) consists of three parts: attention weight $\alpha_{h,i}(t_h)$, temporal decay function $\kappa(t - t_h)$ and historical influence factor $g_{h,i}(t_h)$ between vertex $h$ and $j$. The historical influence factor indicates the degree to which a historical neighbor excites the current event, and can be calculated using the same similarity measure, i.e., $\alpha_{h} = - \Vert{{\bf{u}}^{t_h}_h - {\bf{u}}^{t_h}_j} \Vert ^2$. The temporal decay function $\kappa(t - t_h)$ is a kernel function (e.g. exponential, power-law or Rayleigh kernel) illustrating how the impact of an event fades away over time. Using an exponential kernel \cite{lima2018hawkes}, we have:


\begin{equation}
\kappa(t - t_h) = exp(-\delta_l(t - t_h))
\end{equation}  
where $t_h$ is the time when the historical event occurred, $\delta_l > 0 $ is a vertex-dependent and trainable parameter representing the decay rate for each individual vertex. 
The attention weight $\alpha_{h,i}(t_h)$ models the distinct excitation effects from different historical neighbors. Note that different neighbors may have different impact on a new edge establishment. For example, say both vertices $B$ and $C$ are friends of $A$, but if $B$ is a closer friend, $A$ and $B$ may have more interactions. 
In our deep Hawkes process, we employ the self-attention mechanism \cite{cheng2016long} to model the distinct excitation effects as follow:

\begin{equation}
\tilde{\alpha}_{h,i}(t_h) = {\bf{z}}^\mathrm{T}Relu({\bf{W}}|{\bf{u}}_i^{t_h} - {\bf{u}}_h^{t_h}|)
\end{equation}  
\begin{equation}
\tilde{\alpha}_{h,i}(t_h) = \frac{\tilde{\alpha}_{h,i}(t_h)}{\sum_{\mathcal{H}_i(t_h)} \tilde{\alpha}_{h,i}(t_h)}
\end{equation}  

\noindent where ${\bf{z}} \in \mathbb{R}^{n}$ and  ${\bf{W}} \in \mathbb{R}^{n \times n}$ are weight matrices, $\mathcal{H}_i(t_h)$ is vertex $i$'s historical neighbor set at time $t_h$, and $Relu$ is the rectified linear unit activation function. 

Since the conditional intensity represents the arrival rate of events, its value should always be positive. However, since node similarity is measured by the negative squared Euclidean distance in (4), it is possible for $\tilde{\lambda}$ to be negative. Thus, we further apply a non-linear transfer function $f: \mathbb{R} \rightarrow \mathbb{R}_+$ (i.e., exponential function) to ensure positive intensity for any event, i.e.,

\begin{equation}
\lambda_{i,j}(t) = f(\tilde{\lambda}_{i,j}(t))
\end{equation}  

By modeling the edge formation process based on Hawkes process, we can infer the probability of a vertex establishing an edge during the following time slot. Specifically, for vertex $i$ at time $t$ and its historical neighbor $j$, the probability of forming an edge between $i$ and $j$ can be defined according to their conditional intensity as

\begin{equation}
p(j|i, \mathcal{H}_i(t)) = \frac{\lambda_{i,j}(t)}{\sum_{j'\in{V}}\lambda_{i,j'}(t)}
\end{equation}  
Then, we can minimize the following log likelihood function to capture the temporal dynamics in discrete networks:

\begin{equation}
\mathcal{L}_{DHP}^t = \sum_{(i, j, t) \in {E_t} } log(p(j|i, \mathcal{H}_i(t)))
\end{equation}  

\subsubsection{\bf{Structural Homophily}} Our DHPrep is also capable of extracting the structural information of a network to formulate the closeness or similarity between two vertices such that both spatial and temporal connections are captured for dynamic network representation. Specifically, the first order proximity is adopted in our DHPrep model to capture the current network structural information. Similar to Deep Hawkes process, we utilize Euclidean distance to denote the similarity of a vertex pair, and the loss function is defined as follows: 

\begin{equation}
\mathcal{L}_{1st}^t =  \sum_{i, j = 1}^{N}w_{i,j}^t \Vert {\bf{u}}^{t}_i - {\bf{u}}^{t}_j \Vert ^2
\end{equation}

\noindent The idea of Equation (11) originates from Laplacian Eigenmaps \cite{belkin2003laplacian}, which incurs a penalty when similar vertexes are mapped far away in latent space. As a result, the network structural information can be captured by structural homophily, which is combined with Deep Hawkes process to represent the evolving network dynamics.

\noindent \subsubsection{\bf{Temporal Smoothness}} For temporal network representation, it is desirable to learn the representation smoothly (i.e., avoid abrupt transitions) for different network snapshots because such consistency makes the evolution tendency better observed. Therefore, we apply temporal smoothness \cite{zhou2018dynamic} in our DHPrep to avoid rebuilding the latent representations for snapshots at different times. In other words, it minimizes the Euclidean distance between representation vectors in adjacent snapshots. Accordingly, the loss function can be re-written as:

\begin{equation}
\mathcal{L}_{smooth}^{t}=
\begin{cases}
\sum_{i=1}^{N} \Vert{{\bf{u}}^{t}_i - {\bf{u}}^{t-1}_i} \Vert ^2& t>1\\
0& t=1
\end{cases}
\end{equation}


Based on the aforementioned deep Hawkes process, structural homophily and temporal smoothness techniques, we weave them together into a unified framework, and propose a semi-supervised learning model for dynamic network representations by jointly minimizing the following objective function:

\begin{equation}
\mathcal{L}_{mix} = \sum_{t=1}^{T} \mathcal{L}_{1st}^t + \beta_0 \mathcal{L}_{DHP}^{t} + \beta_1 \mathcal{L}_{smooth}^t
\end{equation}

\noindent where $\beta_0$ and $\beta_1$ are two hyper parameters to trade off different loss functions. Note that the structural homophily loss $\mathcal{L}_{1st}$ and smoothness loss $\mathcal{L}_{smooth}$ are non-linear least square functions, so they can be optimized through gradient descent method. However, optimizing the Deep Hakwes process loss $\mathcal{L}_{DHP}^{t}$ is computationally expensive as it requires the summation over the entire vertex set to calculate the condition intensity $p(\cdot|\cdot)$. Fortunately, due to the exponential transfer function applied in Eq. 6, $p(\cdot|\cdot)$ can be regarded as a $Softmax$ function applied to $\lambda_{i,j}(t)$. In this case, the Negative Sampling method \cite{mikolov2013distributed} can be utilized to approximately optimize the loss for Deep Hawkes Process. Specifically, given an edge $(i, j, t)$, we sample those negative vertices that are not connected to vertex $i$ at time $t$ (i.e., $(i,k,t) \notin{E^t}$), where the probability for a vertex being sampled is proportion to its vertex degree distribution. By obtaining negative samples, the loss term $\mathcal{L}_{DHP}(t)$ can be rewritten as:

\begin{equation}
-\sum_{(i,j,t) \in {E^t}} \bigg ( \text{log} \sigma(\tilde{\lambda}_{ij}(t)) + \sum_{k =1}^K \mathbb{E}_{k}[\text{log} \sigma(-\tilde{\lambda}_{ik}(t))] \bigg)  
\end{equation}

\noindent where $K$ is the number of negative samples and $\sigma(x) = 1/(1+\text{exp}(x))$ is the sigmoid function. Additionally, the computation complexity of $\mathcal{L}_{DHP}^{t}$ is also affected by the time length of the historical neighbors. In our training procedure, we cap the number of the most recent historical snapshots that are used to obtain $\lambda_{ij}(t)$. In particular, Stochastic Gradient Descent (SGD) with Pytorch implementation \cite{goodfellow2016deep} is utilized to optimize our DHPrep model, and in each epoch we sample a
mini-batch of edges (along with their historical neighbors) to update the vertices representations.

\section{Experiments}
In this section, we conduct experiments to evaluate and compare the performance of our proposed DHPrep with state-of-the-art benchmarks. 

\subsection{Datasets} 
To evaluate the effectiveness of DHPrep, we conduct extensive experiments on four real-world datasets. They are all social network datasets where vertices represent people and links indicate interactions among them. Meanwhile, every edge is associated with the time (i.e., date) of its creation so the entire dynamic network can be divided into time series sub-networks. Table I provides the basic statistics of the four datasets. The following is a detailed description of these datasets:

\begin{table}[]
\small
\centering
\caption{Statistics of each dataset.}
\begin{tabular}{lcccc}
\hline
\specialrule{0em}{1pt}{1pt}
Attribute        & \multicolumn{1}{l}{Radoslaw} & \multicolumn{1}{l}{Facebook} & \multicolumn{1}{l}{Calls} & \multicolumn{1}{l}{Mathoverflow} \\ \hline
\specialrule{0em}{1pt}{1pt}
\ Vertices   & 167                          & 1899                         & 6809                      & 16,836                           \\
\specialrule{0em}{1pt}{1pt}
\ Edges      & 82,927                       & 59,835                       & 52,050                    & 203,639     \\
\specialrule{0em}{1pt}{1pt}
\ Time steps & 26                           & 27                           & 35                        & 26                               \\ \hline
\specialrule{0em}{1pt}{1pt}
\end{tabular}
\end{table}

\begin{itemize}
    \item Radoslaw \cite{emailsRadoslaw}. This is an internal email communication network between employees of a mid-sized manufacturing company\footnotemark. It consists of 82,927 email logs with 167 users over 9 months. In this dataset, users are considered as vertices and timesteps are defined as consecutive non-overlapping 10-day periods. During each timestep, if one user emails another, an edge is created and its weight is determined by the number of emails sent between them.
	
	\item Calls \cite{eagle2006reality}. This is a sparse mobile network consisting of 52,048 call logs among 6809 users over 90 days\footnotemark[\value{footnote}]. The time series network is constructed in the same way as Mobile, where each network collects the call logs between a 3-day period, and the weight is determined by the number of calls made between two vertices.

	\item Facebook \cite{opsahl2009clustering}. The Facebook-like Forum Network was attained from an online community of students at University of California, Irvine, in 2004\footnotemark[\value{footnote}]. It has a similar structure with Radoslaw dataset, and includes users that sent or received at least one message. A total number of 59,835 online messages were made among 1899 users. We construct the network in the same way as Calls but with a one-week timestep.
	
	\footnotetext{http://networkrepository.com/}

	\item Mathoverflow \cite{paranjape2017motifs}. This is a temporal network collected by SNAP\footnote{https://snap.stanford.edu/data/sx-mathoverflow.html}, which records the interactions between users on MathOverflow site. It consists 16,836 users with 203,639 temporal interactions among them. The total duration of this dataset is 2350 days, and the timestep is set as 4-month to generate the time series network. 
\end{itemize}

\subsection{Baselines Methods} We compare DHPrep with several state-of-the-art methods (i.e., LINE, Node2vec, Dytriad, TNE, DyRNN and DyAERNN) using their published implementations. The baseline methods are summarized as follows.

\begin{itemize}
    \item LINE \cite{tang2015line}: LINE is a static embedding method which aims to capture first- and second-order proximity. In our experiment, we employ second-order proximity to learn the representations, with negative sample $K = 5$ and total sample $N = 100$ million.
	
	\item Node2vec \cite{grover2016node2vec}: Node2vec is a static embedding method that preserves the network structural information by leveraging random walks, where parameters $p$ and $q$ indicate ‘walk length’ and ‘window size’. In our experiment, all combinations of hyper parameters are tested for $p \in \{0.25, 0.5, 1\}$ and $q \in \{0.25, 0.5, 1\}$

	\item Dytriad \cite{zhou2018dynamic}: Dytriad is a dynamic network representation network which aims to learn discrete network representations, where the temporal dynamics are captured by a triadic closure process. In \cite{zhou2018dynamic},$\beta_0$ and $\beta_1$ are two hyper parameters representing the balance ratio for loss. In our experiments, we set $\beta_0$ and $\beta_1$ with a grid search from $\{0.01, 0.1\}$.
	
	\item TNE \cite{zhu2016scalable}: TNE is a dynamic network embedding approach based on matrix factorization. We obtain the optimal $\lambda$ using a grid search from set $\{0.001, 0.01, 0.1, 1\}$.
	
	\item DyRNN \cite{goyal2020dyngraph2vec}: DyRNN is a deep neural network based dynamic embedding method. It takes network adjacent matrices as input and apply Recurrent Neural Networks to learn representations. Hyper parameter $\beta$ controls the reconstruction weight of non-zero elements. In our experiment, we obtain the best $\beta$ by a grid search from set $\{1, 5, 10\}$
	
	\item DyAERNN \cite{goyal2020dyngraph2vec}: DyAERNN is a variant of DyRNN, where an autoencoder is applied before adjacent matrices are applied to RNN. We tune the hyper parameter $\beta$ in the same way as DyRNN.
\end{itemize}

In our experiments, we first use different methods to obtain embedded vectors of vertices. Then, we gather all samples from different timesteps. Using a Logisitic Regression model as classifier, we repeat the 5-fold cross validation on the gathered sample set for 10 times, and compare the average performance. For LINE, Node2vec and TNE, we take the inner product between vertices representations to indicate their similarity. This is because the optimization objectives for these algorithms are inner product based. For Dytriad, DyRNN, DyAERNN and our DHPrep, we utilize the negative squared Euclidean distance as the similarity indicator. For fair comparisons, we set the embedding dimension $d = 128$, and perform a grid search for all methods to obtain their beset performance.

\subsection{Tasks and Evaluation Metrics} In our experiments, we first learn vertices representations ${\bf{u}}$ at different time steps of the dynamic network $\{G_1, ... , G_T \}$ according to our DHPrep model. Then the learned representation will be applied to the following tasks to evaluate the performance:

\begin{itemize}
    \item Link prediction. This task aims to determine if there will be an edge between two given vertices in the next time step $t + 1$ based on the absolute difference between their representations in the current time step $t$, i.e., $|{\bf{u}}^{t}_i - {\bf{u}}^{t}_j|$.
    
	\item New link prediction. New links are those links occurred during the current time step (i.e., present in the current snapshot but not in previous snapshots). While new links only account for a small percentage of all links, they reveal the true evolution of dynamic networks. It must be emphasized that this task differs from the task of Link Prediction in that we only consider newly added links in the training and test process and ignore preexisting links.
	
	\item Vertices recommendation. For a given vertex, this task aims to select the top-$k$ candidate vertices for recommendation. This task is similar to link prediction, but it considers the ranking order and derives the top-$k$ vertices to evaluate the performance of learned representations. 
\end{itemize}

For link prediction and new link prediction tasks, we adopt F1 and AUC scores as performance metrics. For vertices recommendation, we utilize Precision@$k$ and Recall@$k$ as performance metrics. In addition to these tasks, we further provide parameter sensitivity studies (i,e., analyze the historical influence by using different decay kernels and different length of historical snapshots). 

\begin{table*}[]
\small
\centering
\caption{Comparison of DHPrep with state-of-the-art approaches in link prediction task}
\begin{tabular}{lllllllll}
\hline
\specialrule{0em}{1pt}{1pt}
\multirow{2}{*}{Methods} & \multicolumn{2}{c}{Radoslaw}                     & \multicolumn{2}{c}{Facebook}                     & \multicolumn{2}{c}{Calls}                        & \multicolumn{2}{c}{Mathoverflow}                 \\ \cline{2-9} 
                         \specialrule{0em}{1pt}{1pt}
                         & \multicolumn{1}{c}{F1} & \multicolumn{1}{c}{AUC} & \multicolumn{1}{c}{F1} & \multicolumn{1}{c}{AUC} & \multicolumn{1}{c}{F1} & \multicolumn{1}{c}{AUC} & \multicolumn{1}{c}{F1} & \multicolumn{1}{c}{AUC} \\ \hline
\specialrule{0em}{1pt}{1pt}
Node2vec                 & 66.01$\pm$0.19         & 62.65$\pm$0.18         & 61.79$\pm$0.43          & 59.50$\pm$0.51          & \textbf{81.25}$\pm$\textbf{0.16} & \textbf{81.24}$\pm$\textbf{0.20} & 56.46$\pm$0.26          & 54.82$\pm$0.25          \\
\specialrule{0em}{1pt}{1pt}
LINE                     & 67.90$\pm$0.37         & 65.94$\pm$0.37          & 62.53$\pm$0.12          & 61.69$\pm$0.19          & 68.12$\pm$0.35          & 68.43$\pm$0.31          & 55.34$\pm$0.19          & 56.44$\pm$0.19          \\
\specialrule{0em}{1pt}{1pt}
DHPrep                   & \textbf{72.00}$\pm$\textbf{0.28} & \textbf{70.44}$\pm$\textbf{0.27} & \textbf{71.14}$\pm$\textbf{0.41} & \textbf{71.95}$\pm$\textbf{0.31} & 80.18$\pm$0.27        & 81.20$\pm$0.23         & \textbf{71.44}$\pm$\textbf{0.09} & \textbf{73.31}$\pm$\textbf{0.06} \\ \hline
\specialrule{0em}{1pt}{1pt}
TNE                      & 71.66$\pm$0.37          & 71.72$\pm$0.28          & 51.72$\pm$0.68            & 53.02$\pm$0.53           & \textbf{77.45}$\pm$\textbf{0.21} & \textbf{77.44}$\pm$\textbf{0.13} & 52.33$\pm$0.20          & 53.93$\pm$0.15          \\
\specialrule{0em}{1pt}{1pt}
DyAE                     & 49.08$\pm$0.23          & 55.69$\pm$0.11          & 45.19$\pm$0.97            & 55.98$\pm$0.47           & 57.46$\pm$0.52          & 61.17$\pm$0.41          & 39.54$\pm$0.52          & 54.96$\pm$0.98          \\
\specialrule{0em}{1pt}{1pt}
DyAERNN                  & 63.25$\pm$0.14          & 61.72$\pm$0.05          & 51.61$\pm$0.26            & 57.21$\pm$0.33           & 59.80$\pm$0.41          & 60.97$\pm$0.21          & 55.93$\pm$0.11          & 58.72$\pm$0.10          \\
\specialrule{0em}{1pt}{1pt}
Dytriad                  & \textbf{77.30}$\pm$\textbf{0.14}          & \textbf{78.17}$\pm$\textbf{0.47}         & 60.58$\pm$0.58            & 60.77$\pm$0.51           & 68.84$\pm$0.30          & 69.86$\pm$0.32          & 64.30$\pm$0.09          & 65.51$\pm$0.05          \\
\specialrule{0em}{1pt}{1pt}
DHPrep                   & 73.55$\pm$0.45 & 71.20$\pm$0.53 & \textbf{72.97}$\pm$\textbf{0.24}   & \textbf{74.25}$\pm$\textbf{0.31}  & 73.74$\pm$0.34          & 73.07$\pm$0.35          & \textbf{67.65}$\pm$\textbf{0.12} & \textbf{69.33}$\pm$\textbf{0.09} \\ \hline
\end{tabular}
\end{table*}

\subsection{Experimental Results} We evaluate the quality of learned representations by feeding them into the aforementioned tasks.

\textbf{Link Prediction} In order to perform link prediction, for each dataset we gather all links and randomly sample a fixed number of non-links (unconnected vertex pairs) from all network snapshots. Then, we calculate features of all samples according to their vertices representations. Based on these features, we use a Logisitic Regression model as classifier and repeat 5-fold cross validation on the gathered sample set for 10 times to get the average link prediction results. We use F1 and AUC score as performance metrics. Note that all static network representation approaches only consider those vertices that have edges in the current time step, and thus cannot learn representations for those vertices presented previously. Therefore, we compare static and dynamic approaches separately. 

Table II compares the experimental results obtained from the aforementioned methods. For static approaches, our DHPrep outperforms other methods in most cases, which shows the effectiveness of our method that takes advantage of the inherent temporal dynamics of the networks. Note that there is an exception in Calls dataset, where Node2vec algorithm performs slightly better than ours. The reason is that Calls dataset is sparse (i.e., high vertex-to-edge ratio) and the network structure is quite static. In this case, the current time step representations for Node2vec achieves similar or even better performance on link prediction. For dynamic approaches, our DHPrep achieves the best performance on Facebook and Mathoverflow datasets. However, Dytraid shows better prediction result on Radoslaw dataset. A possible reason is that the user relationship doesn't last for long so the connection pattern changes a lot over time. In particular, when the network state transition follows a Markov random process, it brings much noise to DHPrep for link prediction. For Calls dataset, TNE and DHPrep have the best and the second-best predictions. This is because TNE aims to minimize the sudden change of representations, making it suitable for networks with small temporal dynamics. Overall, our DHPrep better models the network evolution and achieves the best performance in link predictions. 

\begin{table*}[]
\small
\centering
\caption{Comparison of DHPrep with state-of-the-art approaches in new link prediction task}
\begin{tabular}{lllllllll}
\hline
\specialrule{0em}{1pt}{1pt}
\multirow{2}{*}{Methods} & \multicolumn{2}{c}{Radoslaw}                     & \multicolumn{2}{c}{Facebook}                     & \multicolumn{2}{c}{Calls}                        & \multicolumn{2}{c}{Mathoverflow}                 \\ \cline{2-9} 
                         \specialrule{0em}{1pt}{1pt}
                         & \multicolumn{1}{c}{F1} & \multicolumn{1}{c}{AUC} & \multicolumn{1}{c}{F1} & \multicolumn{1}{c}{AUC} & \multicolumn{1}{c}{F1} & \multicolumn{1}{c}{AUC} & \multicolumn{1}{c}{F1} & \multicolumn{1}{c}{AUC} \\ \hline
\specialrule{0em}{1pt}{1pt}
Node2vec                 & 56.29$\pm$1.44         & 55.00$\pm$1.49         & 52.33$\pm$1.02          & 52.02$\pm$0.86          & 61.87$\pm$0.16         & 62.42$\pm$2.03        & 56.68$\pm$0.37          & 58.35$\pm$0.26          \\
\specialrule{0em}{1pt}{1pt}
LINE                     & 62.23$\pm$1.30         & 60.88$\pm$1.21         & 61.94$\pm$0.37          & 61.66$\pm$0.19          & 67.30$\pm$1.18         & 67.41$\pm$1.24        & 68.92$\pm$0.17         & 68.58$\pm$0.15          \\
\specialrule{0em}{1pt}{1pt}
DHPrep                   & \textbf{64.16}$\pm$\textbf{0.50} & \textbf{62.74}$\pm$\textbf{0.42} & \textbf{67.98}$\pm$\textbf{1.34} & \textbf{69.77}$\pm$\textbf{1.20} & \textbf{77.81}$\pm$\textbf{1.119}        & \textbf{77.94}$\pm$\textbf{0.88}         & \textbf{69.80}$\pm$\textbf{0.89} & \textbf{70.65}$\pm$\textbf{0.18} \\ \hline
\specialrule{0em}{1pt}{1pt}
TNE                      & 52.73$\pm$0.25         & 53.11$\pm$0.26         & 71.39$\pm$0.52           & 70.49$\pm$0.48         & 53.71$\pm$0.11         & 54.59$\pm$0.11        & 66.95$\pm$0.43          & 64.74$\pm$0.30          \\
\specialrule{0em}{1pt}{1pt}
DyAE                     & 44.74$\pm$0.66         & 55.67$\pm$0.24         & 60.04$\pm$0.78           & 63.08$\pm$0.48         & 32.34$\pm$3.51         & 54.54$\pm$0.63        & 48.29$\pm$1.10          & 54.88$\pm$0.63           \\
\specialrule{0em}{1pt}{1pt}
DyAERNN                  & 52.18$\pm$0.51         & 56.76$\pm$0.20         & 63.76$\pm$0.61           & 64.39$\pm$0.41         & 57.89$\pm$0.16         & 59.81$\pm$0.11        & 58.22$\pm$0.51          & 57.90$\pm$0.68          \\
\specialrule{0em}{1pt}{1pt}
Dytriad                  & 57.59$\pm$0.28         & 57.62$\pm$0.21         & 65.78$\pm$0.32           & 66.99$\pm$0.32         & 62.84$\pm$0.32         & 63.73$\pm$0.27        & 62.90$\pm$0.77          & 63.35$\pm$0.69          \\
\specialrule{0em}{1pt}{1pt}
DHPrep                   & \textbf{71.55}$\pm$\textbf{0.34} & \textbf{72.92}$\pm$\textbf{0.24} & \textbf{73.97}$\pm$\textbf{0.60}   & \textbf{71.74}$\pm$\textbf{0.63}  & \textbf{67.37}$\pm$\textbf{0.12}          & \textbf{68.70}$\pm$\textbf{0.11}          & \textbf{69.39}$\pm$\textbf{0.39} & \textbf{66.98}$\pm$\textbf{0.43} \\ \hline
\end{tabular}
\end{table*}

\textbf{New Link Prediction} In dynamic networks, fresh links dominate the networks evolution. In many applications, new links representing new interactions or relationship usually contain more information and thus deserve more attention. For example, in a co-author network, a new link between two scholars indicates their new collaboration, which may suggest the change of their research interests. Therefore, we further compare our DHPrep with other approaches in new link prediction. Specifically, if a link presents in the current network $G_{t}$ but not in $G_{t-1}$, it is called a new link. Experiment settings are similar to link prediction and the results are shown in Table III.

From Table III, we can observe that our DHPrep algorithm achieves the best new link prediction results on both metrics (F1 and AUC scores) among all (static and dynamic) approaches. This is because our DHPrep utilizes both structural and temporal information to learn network representations. It is worth noting that, for Calls dataset, our DHPrep outperforms Node2vec in new link prediction, indicating the effectiveness of DHPrep in capturing temporal dynamics. For dynamic approaches, DHPrep consistently achieves the best performance. We can also see that Dytriad has the worse performance due to the fact that it only considers adjacent snapshot for modeling temporal dynamics. On the other hand, even though DyAERNN and TNE consider multiple historical snapshots, they cannot reveal the temporal process of the edge formation during the network evolution. 

\begin{figure}
\centering
\subfigure[Precesion@K on facebook]{
\includegraphics[width=4.2cm]{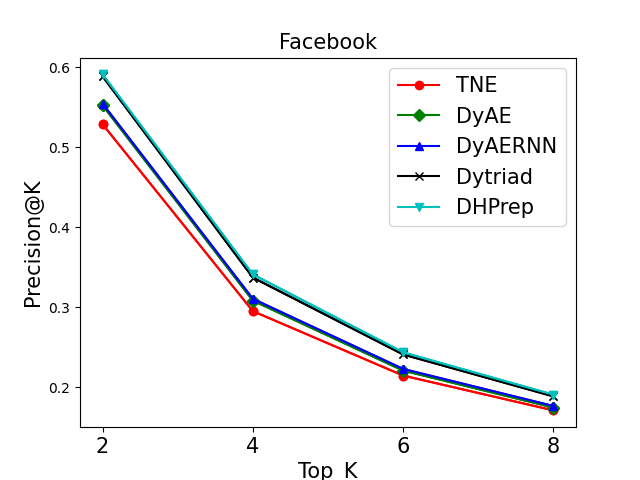}}
\subfigure[Recall@K on facebook]{
\includegraphics[width=4.2cm]{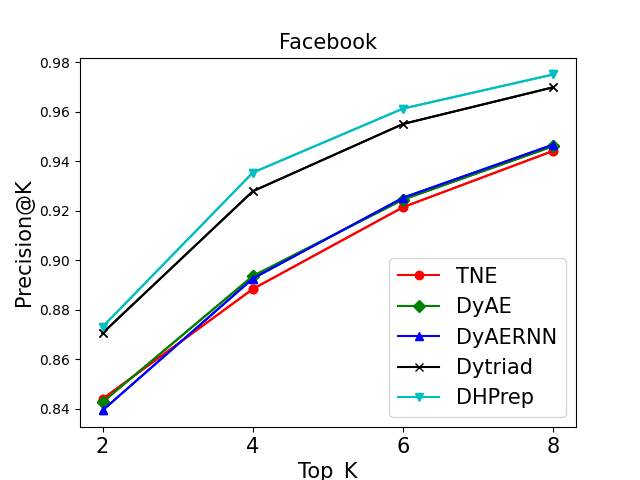}}
\caption{Node Recommendation Analysis on Facebook dataset}
\end{figure}

\begin{figure}
\centering
\subfigure[Precesion@K on Mathoverflow]{
\includegraphics[width=4.2cm]{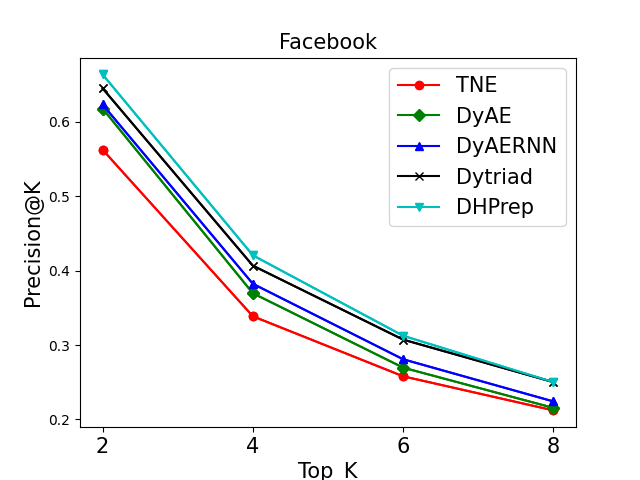}}
\subfigure[Recall@K on Mathoverflow]{
\includegraphics[width=4.2cm]{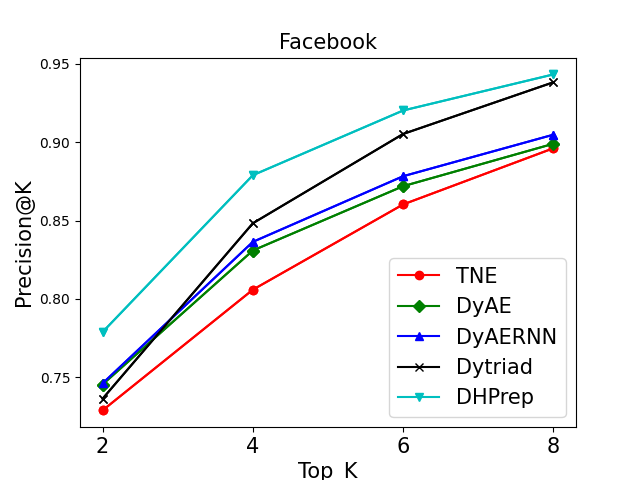}}
\caption{Node Recommendation Analysis on Mathoverflow dataset}
\end{figure}

\textbf{Vertices Recommendation} Beside link prediction, vertices recommendation is another important application for performance evaluation. Specifically, for each vertex $i$ at time step $t$, we predict the top-$K$ possible neighbors to vertex $i$ at next time step $t+1$. The ranking score is based on the linkage probability, which is calculated in the same way as in link prediction setting. Then, we can derive the top-$K$ vertices with the highest ranking scores as recommendation candidates.  

Experimental results are shown in Figure 2 and Figure 3. Specifically, Figure 2 shows the Recall@K and Precision@K results using Facebook dataset. Compared with TNE, DyAE, and DyAERNN, on average our DHPrep improves the precision and recall by $7.25\%$ and $14.19\%$ respectively. For Mathoverflow dataset, the results are shown in Figure 3, where our DHPrep on average improves the precision and recall by $19.54\%$ and $18.65\%$ respectively. Among all algorithms, Dytriad is the second best. In terms of precision and recall performance, our DHPrep outperforms Dytriad by $2.14\%$ and $1.2\%$ on Facebook dataset, and $9.21\%$ and $2.35\%$ on Mathoverflow dataset. This results suggest that, by utilizing the triad closure process to model the network dynamics, Dytriad is able to capture the closest relationship between vertices. However, in link prediction task, 
the advantage of DHPrep over Dytriad is quite obvious. Therefore, we can infer that Dytriad has poor performance in recommending estranged vertices. Overall, our DHPrep has the best performance because it uses the temporal point process to model edge formation, and thus can capture temporal dynamics and make the vertex representations more informative. 

\begin{figure*}
\centering
\subfigure[The impact of time decay kernels.]{
\includegraphics[width=3.5in]{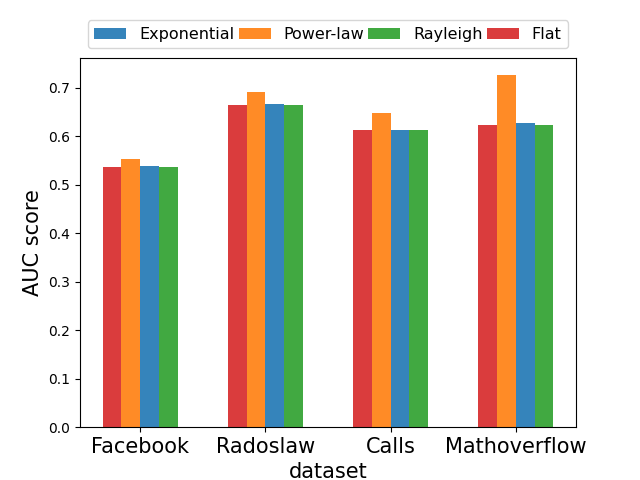}}
\subfigure[The impact of historical networks snapshots]{
\includegraphics[width=3.5in]{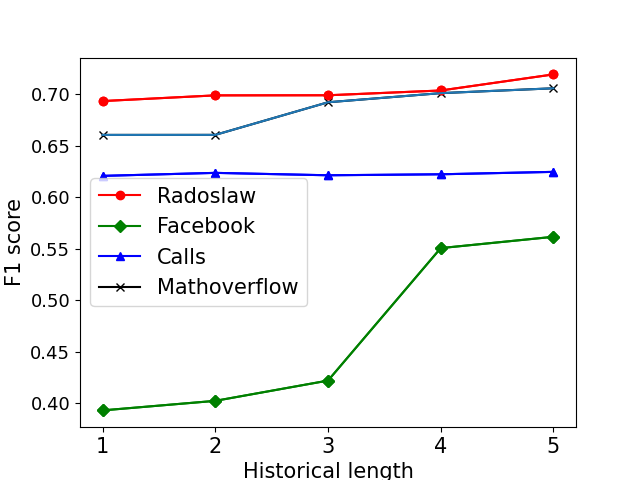}}
\caption{Historical influence analysis}
\end{figure*}

\begin{figure*}
\centering
\subfigure[Precesion@K on Mathoverflow]{
\includegraphics[width=3.5in]{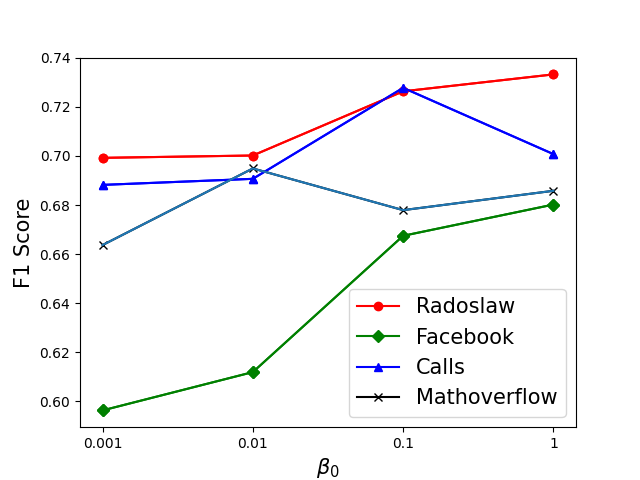}}
\subfigure[Recall@K on Mathoverflow]{
\includegraphics[width=3.5in]{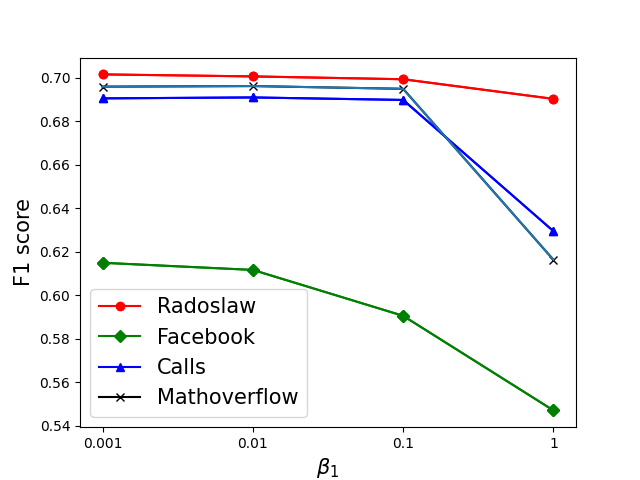}}
\caption{The effect of hyper parameters $\beta_0$ and $\beta_1$}
\end{figure*}

\subsection{Performance w.r.t. Historical influence} 
In order to analyze the historical influence for dynamic networks evolution, we further investigate how time decay kernels and the length of historical snapshots can affect the prediction performance for learned representations.

\textbf{Kernel function analysis} To analyze the time decay effect for historical snapshots, we use four different kernels (exponential, power-law, Rayleigh and flat kernels) for representation learning. Here, the flat kernel means that all historical influences are equal so there is no temporal decay for them. Apparently, the power-law kernel decays slower than the exponential kernel, while the Rayleigh kernel experiences the fastest decay. The experimental result are shown in Figure 4(a). We can see that the Power-law kernel achieves the best prediction performance. This is because, with the Power-law kernel, the historical influence has the slowest temporal decay so that it can effectively capture the temporal dynamics. On the other hand, the flat kernel achieves the worst performance because it cannot differentiate temporal information from different historical network snapshots. These results further proof the effectiveness of our proposed DHPrep algorithm in capturing the temporal dynamics for dynamic network representations.

\textbf{Historical length analysis} In order to analyze historical influences, we further study the effect of the length of historical snapshots used to model temporal dynamics. The results are shown in Figure 4(b). We can see that DHPrep algorithm achieves the highest F1 score when five historical snapshots are used. The more historical snapshots DHPre uses, the better performance it can achieve. Since the historical information models the evolution of dynamic networks, the number of historical snapshots plays a key role in DHPrep in determining the performance of learned representations.

\subsection{Parameter Sensitivity}
We investigate the parameter sensitivity in this sub-section. In our DHPrep model, there are two hyper parameters $\beta_0$ and $\beta_1$ in the loss function, where $\beta_0$ indicates the weight of deep Hawkes process in Equation (13), and $\beta_1$ represents the weight of temporal smoothness. We study the sensitivity of each parameter by fixing the other one. Figure 5 shows the link prediction results on four datasets with respect to $\beta_0$ and $\beta_1$. Note that $\beta_0$ defines the importance of deep Hawkes process that models the network evolution process. From Figure 5(a), we can see that the F1 score increases with $\beta_0$ in most cases. However, there are exceptions (in Calls dataset and Mathoverflow dataset) that larger $\beta_0$ leads to worse performance. Therefore, it is important to set the value of $\beta_0$ properly. From Figure 5(b), we observe that the performance degrades with $\beta_1$ in all cases. This is because a larger $\beta_1$ limits the network evolution by imposing more penalty when the representation changes drastically.

\section{Conclusion}
In this paper, we present a novel semi-supervised representation learning algorithm, named DHPrep, for discrete dynamic networks. Specifically, DHPrep incorporates structural information and Hawkes process based temporal information to model the network evolution pattern over time, so that the learned vertices representations can capture the temporal dynamics of the network. In order to evaluate the effectiveness of DHPrep, we conduct comprehensive experiments on four real-world datasets. Experimental results demonstrate that DHPrep outperforms baseline approaches in various tasks. Our future work is to generalize DHPrep by incorporating branching process to differentiate influences from different neighbors.

\bibliographystyle{IEEEtran}

\bibliography{paper}

\end{document}